\documentclass[12pt,table]{article}

\usepackage{scicite}
\usepackage{times}
\usepackage{graphicx}
\usepackage{booktabs}
\usepackage{multirow}
\usepackage[normalem]{ulem}
\usepackage{amsmath}
\usepackage{float}
\usepackage{tikz}
\usetikzlibrary{arrows}
\usepackage{url}
\usepackage[font=footnotesize]{subcaption}
\usepackage{placeins}   		
\useunder{\uline}{\ul}{}
\usepackage{lscape}
\usepackage{longtable}
\usepackage{glossaries}
\usepackage{makecell}
\usepackage{xcolor}

\newacronym{SI}{SI}{Supplementary Information}

\newenvironment{sciabstract}{%
\begin{quote} \bf}
{\end{quote}}

\title{Disadvantaged Communities Have Lower Access to Urban Infrastructure}

\author
{Leonardo Nicoletti$^{1\ast}$, Mikhail Sirenko$^{1}$, Trivik Verma$^{1}$\\
\\
\normalsize{$^{1}$Faculty of Technology, Policy and Management,} \\
\normalsize{Delft University of Technology, 2628BX Delft, The Netherlands}\\
\\
\normalsize{$^\ast$To whom correspondence should be addressed; E-mail: t.verma@tudelft.nl}
}

\date{}

\begin{document}

\baselineskip24pt

\maketitle



\begin{sciabstract}
Disparity in spatial accessibility is strongly associated with growing inequalities among urban communities. Since improving levels of accessibility for certain communities can provide them with upward social mobility and address social exclusion and inequalities in cities, it is important to understand the nature and distribution of spatial accessibility among urban communities. To support decision-makers in achieving inclusion and fairness in policy interventions in cities, we present an open-source and data-driven framework to understand the spatial nature of accessibility to infrastructure among the different demographics. We find that accessibility to a wide range of infrastructure in any city ($54$ cities) converges to a Zipf's law, suggesting that inequalities also appear proportional to growth processes in these cities. Then, assessing spatial inequalities among the socioeconomically clustered urban profiles for $10$ of those cities, we find urban communities are distinctly segregated along social and spatial lines. We find low accessibility scores for populations who have a larger share of minorities, earn less, and have a relatively lower number of individuals with a university degree. These findings suggest that the reproducible framework we propose may be instrumental in understanding processes leading to spatial inequalities and in supporting cities to devise targeted measures for addressing inequalities for certain underprivileged communities.

\end{sciabstract}




\section*{Introduction}


Cities are places of consumption that provide a variety of amenities and services to their citizens \cite{jayne2005cities}. Cities with high-capital investments and technological dominance have become attractive for migrants who are highly skilled, well-educated, and well-paid \cite{huang2011spatial,scott2017constitution}. As \textit{Nijman et al.} \cite{nijman2020urban} argue in a comprehensive review of urban inequalities; national and regional level policies around globalisation, migration and free-market investments have catapulted the urban economy but the benefits have not accrued equitably across urban populations, and as cities grow, the benefits of living in them are increasingly unequally distributed \cite{heinrich2021scaling}. Akin to the notion that to survive a decline, cities must make attractive amenities accessible for increasingly rich workers \cite{glaeser2001consumer}, the responsibility\footnote{Sustainable Development Goal 10 (\textit{Reduced Inequalities}) suggests that there is a mutual understanding among world governments to tackle inequalities as a global challenge.} \cite{economic2017progress} of \emph{equitably} addressing urban inequalities also lies with municipal authorities \cite{seskin2012complete}.

Over time, the dynamics of gentrification, the housing market, residential segregation and \textit{de jure} practices like exclusionary zoning laws \cite{higginbotham1990jure,parekh2017more} have pushed low-income minority groups to live in deteriorated urban areas receiving a minor share of public investments \cite{rothstein2015racial,rothstein2017color}. Such processes have resulted in the concentration of urban poverty and the creation of strong inequalities among urban communities \cite{hajnal1995nature, musterd2017socioeconomic, sydes2019immigrants, bayon2018place, walks2006ghettos}. Through sociological investigations and observations, scholars have illustrated how urban inequalities are related to interlinked social factors across spatial and temporal scales \cite{nijman2020urban}. In general, depending on their socioeconomic status \cite{mcfarlane2008political}, some communities benefit from access to a wide range of urban services such as mobility, health, education and community space, while others are segregated in neighbourhoods where they struggle to access resources within the city \cite{delafontaine2011impact, tao2014spatial, mayaud2019urban, neutens2015accessibility, van2011discussing}. People living in areas with a poor concentration of accessible services and amenities tend to snowball into low levels of education, poor physical and mental health, disproportionate job opportunities and social exclusion \cite{massey1987effect,gobillon2007effect,glaeser2009inequality,rothstein2017color}, thus lacking upward mobility.

As Eric Klinenberg posits in their book that building better social infrastructure will result in more resilient and less unequal societies, it is an account of how urban infrastructure can facilitate shared spaces through which civic life can be rebuilt and improved \cite{klinenberg2018palaces} for vastly underprivileged communities. Fortunately, globally, cities have implemented policies aimed at improving the quality of urban infrastructure, and at reducing inequalities in accessibility among urban communities \cite{city_of_london, mairie_de_paris_2018, ajuntament, walkable_and_bikeable_cities_lessons_from_seoul_and_singapore_2016, transport_strategy_2030_2019, declaracion_de_bogota}. In the United States and Canada alone, more than $400$ municipalities have implemented policies to encourage the development of ``complete street'' \cite{seskin2012complete}. Such policies aim to create socially inclusive communities with improved \textit{access} to services, shops and recreation, healthy and active lifestyles, more walking and less driving, attractive public spaces, and improved economic vitality \cite{seskin2012complete}.

Even though nuanced variations exist \cite{access-manual,hewko2002measuring, biazzo2019general}, more formally, \emph{Accessibility} refers to the ease with which residents within a city can reach amenities or opportunities (destinations, social interactions, jobs) \cite{hansen1959accessibility}. It is a general measure adopted by planners to understand how land-use and transport systems shape the quality of life of residents in a city and identify where equitable developments can improve the life of marginalised communities \cite{access-manual}. Studies that attempt to inform policy by quantitatively highlighting urban inequalities in cities usually measure variability in access for residents to specific spatial factors, such as access to transit, health or green space infrastructure, and by specific demographic attributes, such as income or ethnicity \cite{jang2017assessing,mayaud2019urban,xiao2017assessment}. These findings are especially useful for decision-makers to tailor policies toward their specific communities \cite{seskin2012complete}. However, inequalities among communities are shaped differently depending on the diversity in distribution of resources and compounded by several other spatial and economic factors \cite{moro2021mobility}, and adopting utilitarian (or case-specific) approaches may prevent in addressing the causes of inequalities in urban regions \cite{farmer2011uneven}. Therefore, systematically understanding the variability in spatial distribution of access and the associated demographic distribution is instrumental in designing targeted and equitable policies for addressing urban inequalities.

Using open-source data, we present a framework to understand and evaluate the differences in access to urban infrastructure afforded to various demographics within cities around the world. By taking into consideration more than $50$ types of urban amenities, we quantify accessibility to different categories of basic services for $54$ cities, globally. Despite the geo-political differences that exist across these cities, we find that accessibility in all cities is driven by a universal log-normal distribution. We infer the existence of this law by comparing the distribution of accessibility to the spread of population density in those regions. By taking into consideration several categories of socioeconomic attributes, we characterize city-specific urban communities and assess inequalities in accessibility in $10$ of those cities (due to data-availability). Our findings reveal the existence of structural inequalities for similar archetypes of urban communities across all North American cities within our data. This approach supports us in investigating the nature of inequalities in access and its relationship with different socioeconomic aspects of urban communities, thus providing evidence to derive tailored policies in addressing inequalities in an equitable manner.

Our goal is twofold: First, we investigate the nature of spatial factors that relate to accessibility in any given city, and in turn urban inequalities. Then, because of its modular nature, the framework can be adapted to the local values of decision-makers with regards to accessibility, and provide support in identifying region-specific social groups suffering from inequalities in access within their regions.

\section*{Results}
\label{sec:results}

\textbf{There is more to the scale-free nature of accessibility \footnote{Subtitle inspired by \cite{cristelli2012there}}}

To examine how accessible urban infrastructure is for residents in a city, we first define a measurement of \emph{accessibility} $A$ (see Section Methods for details) that quantifies the ease with which residents within a city can reach amenities \cite{access-manual,biazzo2019general} (for e.g. Mobility, Active living, Entertainment, Food Choices, Community Space, Education, or Health and Well Being) by walking or by using a form of \emph{active} transportation (e.g. bicycle or roller-blades) \cite{vale2016active}. Active accessibility measures \cite{vale2016active} play a central role in shaping accessibility policy worldwide \cite{seskin2012complete, orozco2019quantifying} by demonstrating how well urban dwellers have access to basic services in the immediate proximity of their place of residence. We make the choice of defining accessibility as \textit{active} for two reasons: First, there are many cities that do not have any reasonable forms of public transportation and presuppose participation in society through private modes of transportation like cars - using a more general definition of accessibility will only facilitate a comparison among limited car-dependent urban regions. Second, our work is responding to policy efforts in promoting more active accessibility in cities around the world \cite{economic2017progress}.

\begin{figure}[H]
\centering
\includegraphics[width = 1\textwidth]{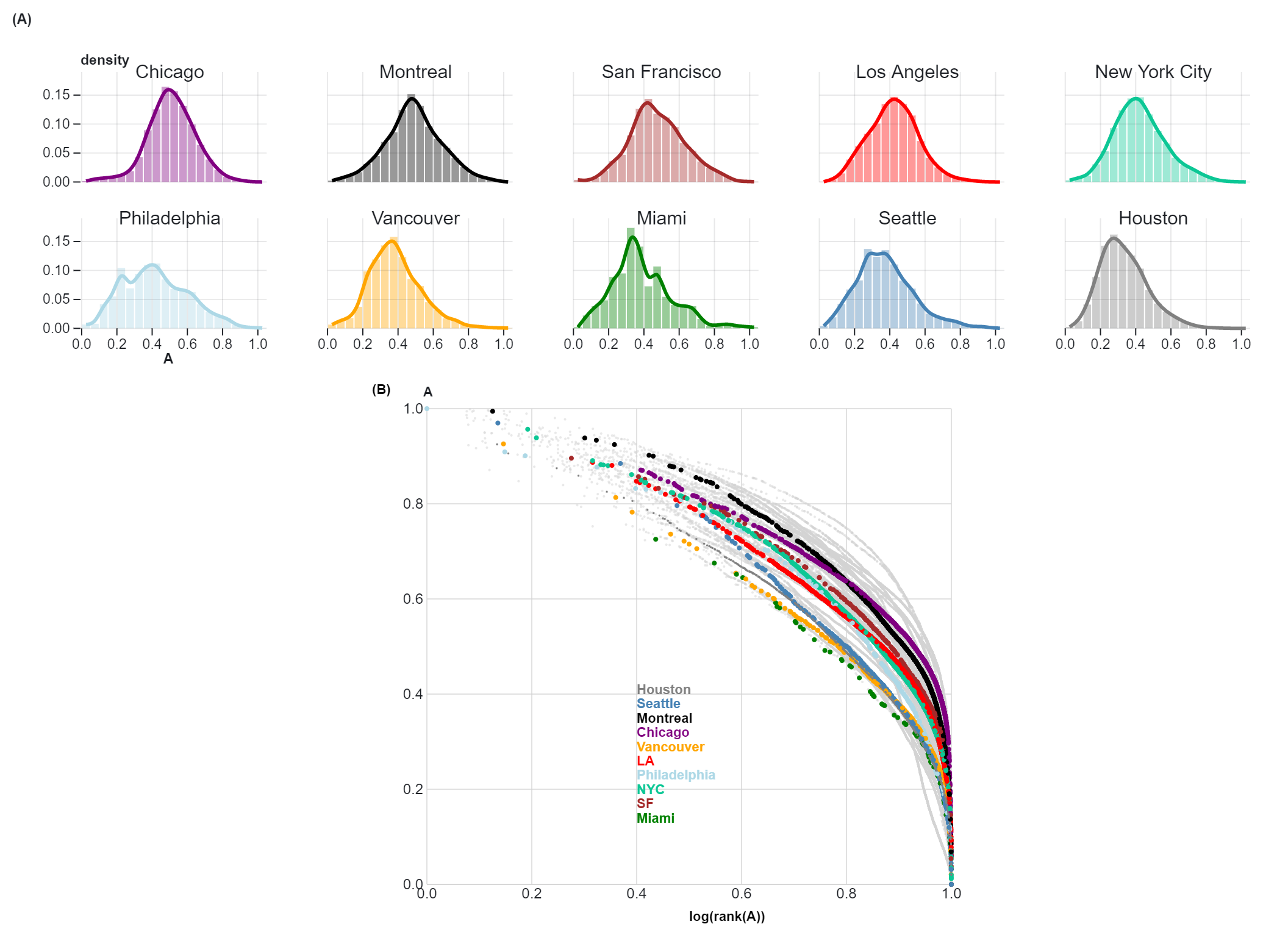}
\caption{(A) Statistical distribution of accessibility for Chicago, Montreal, San Francisco, Los Angeles, New York City, Philadelphia, Vancouver, Miami, Seattle and Houston, showing log-normal behaviour. (B) The scale free relationship between a spatial unit's accessibility A and the rank of its accessibility ($A_i = -0.3 + 1.25(ln(A_i^{r})$). The dots represent equal sized grid-cells within and across all cities.}
\label{fig:fig1}
\end{figure}

Regardless of how each city has evolved, the accessibility score $A \in[0,1]$ for the spatial units (see Section Methods for details) of any given city is either normally distributed (see Figure \ref{fig:fig1}A) or a mixture of normal distributions (see \gls{SI} for a detailed analysis of the distributions). As $A$ is defined using a logarithm, a log-normal relationship governing the distribution of accessibility across the world could be explained following the economic and spatial integration of settlements into a larger urban area \cite{cristelli2012there}: Access (as a resource that is developing with the growth of an urban region) may have evolved log-normally as a result of the overall effects of factors such as economic, technological or demographic changes \cite{parr1973settlement} (also see Ref. \cite{parr1973settlement} for justification of using a log-normal distribution). Though, accessibility score $A_i$ of a given spatial unit $i$ across a city is inversely proportional to the natural log of its rank in accessibility (see Figure \ref{fig:fig1}B) following $A_i = -0.3 + 1.25(ln(A_i^{r})$ where $A_i^{r}$ is the rank of the spatial unit $i$ in the distribution of a specific city, there is a cut-off to the distribution implying that certain regions may not be economically well integrated \cite{cristelli2012there}, depending on the sprawl size of the city \cite{gabaix1999zipf}.

\begin{figure}[H]
\centering
\includegraphics[width = 1\textwidth]{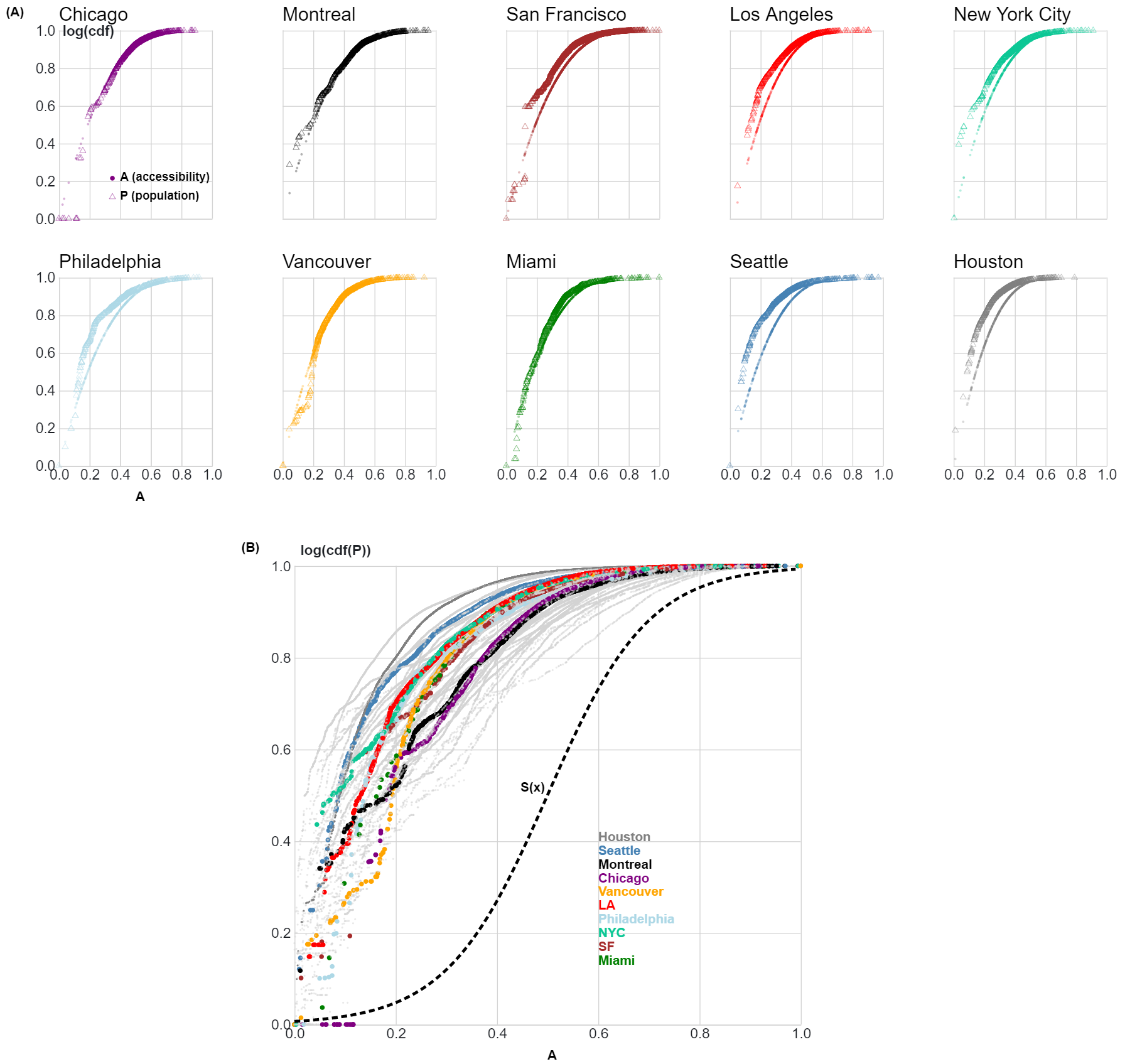}
\caption{(A) Cumulative density of accessibility and cumulative population proportion. On the y-axis the dotted line represents the log of the cumulative density of accessibility, and the triangle symbol line represents the log of the cumulative density of population proportion. The x-axis represents the accessibility score A. (B) Cumulative population proportion versus accessibility score A for the 54 cities studied. $S(x)$, where $S(x) = \frac{1}{1 + e^{-k(x - 0.5)}}$ where $k = 10$, can be used as a benchmark for accessibility within cities, where $50\%$ of the population has at least an accessibility score A of 0.5. The dots and triangles represent equal sized 0.0625 km\textsuperscript{2} spatial units within and across all cities.}
\label{fig:fig2}
\end{figure}

Since accessibility is only defined as a measure of infrastructural provision, decision-makers may argue that regions with lower populations may not necessitate the same investment of resources in providing better levels of accessibility compared to urban centres. To ground the discussion of accessibility in urban policy, we measure the cumulative size of population that has accessibility $A$ lower than $a$ in a city using a separate Global Human Settlement Layer (GHSL) data set of population residing in similar sized spatial units \cite{schiavina2019ghs,freire2016development} (see Section Methods for details about GHSL population grids). Figure \ref{fig:fig2}A shows that, in many North American cities, while there is a large number of spatial units with high accessibility scores (for example, $A > 0.3$), there is a large proportion of the population with low accessibility scores (i.e. $A < 0.3$), evident from the difference in area between the density curves. In Houston and Seattle, for example, around $90\%$ of the population resides in spatial units that have a score $A$ of less than $0.3$. In those cities, accessibility only benefits a few residents. Figure \ref{fig:fig2}B presents this distribution for all $54$ cities within our study. It shows that, in cities where curves are heavily shifted to the right and which start to resemble the sigmoid curve $S(x)$, $S(x) = \frac{1}{1 + e^{-k(x - 0.5)}}$ where $k = 10$), there is a larger share of people that benefits from higher accessibility scores, and the share of people with lower accessibility scores is smaller. Cities like Zurich, for example, come close, but no city resembles the sigmoid function. In this context, $S(x)$ can be used as a benchmark for accessibility within cities, a hypothetical scenario where $50\%$ of the population has at least an accessibility score $A$ of $0.5$. See \gls{SI} for more details on descriptive statistics of accessibility for all cities.

Through this analysis, we are not attempting to resolve the regionalisation debate \cite{cristelli2012there}, wherein, depending on the number of spatial units taken into account, the distribution of $A$ may transform. Our data represents administrative units of each city, which may differ from economically functional units, especially for smaller cities \cite{dijkstra2019eu} (see Section Methods for data description). To partly address this concern, we use a cumulative function to account for sampling biases in low access regions \cite{newman2005power} in illustrations in Figure \ref{fig:fig2}.

\begin{figure}[H]
\centering
\includegraphics[width = 1\textwidth]{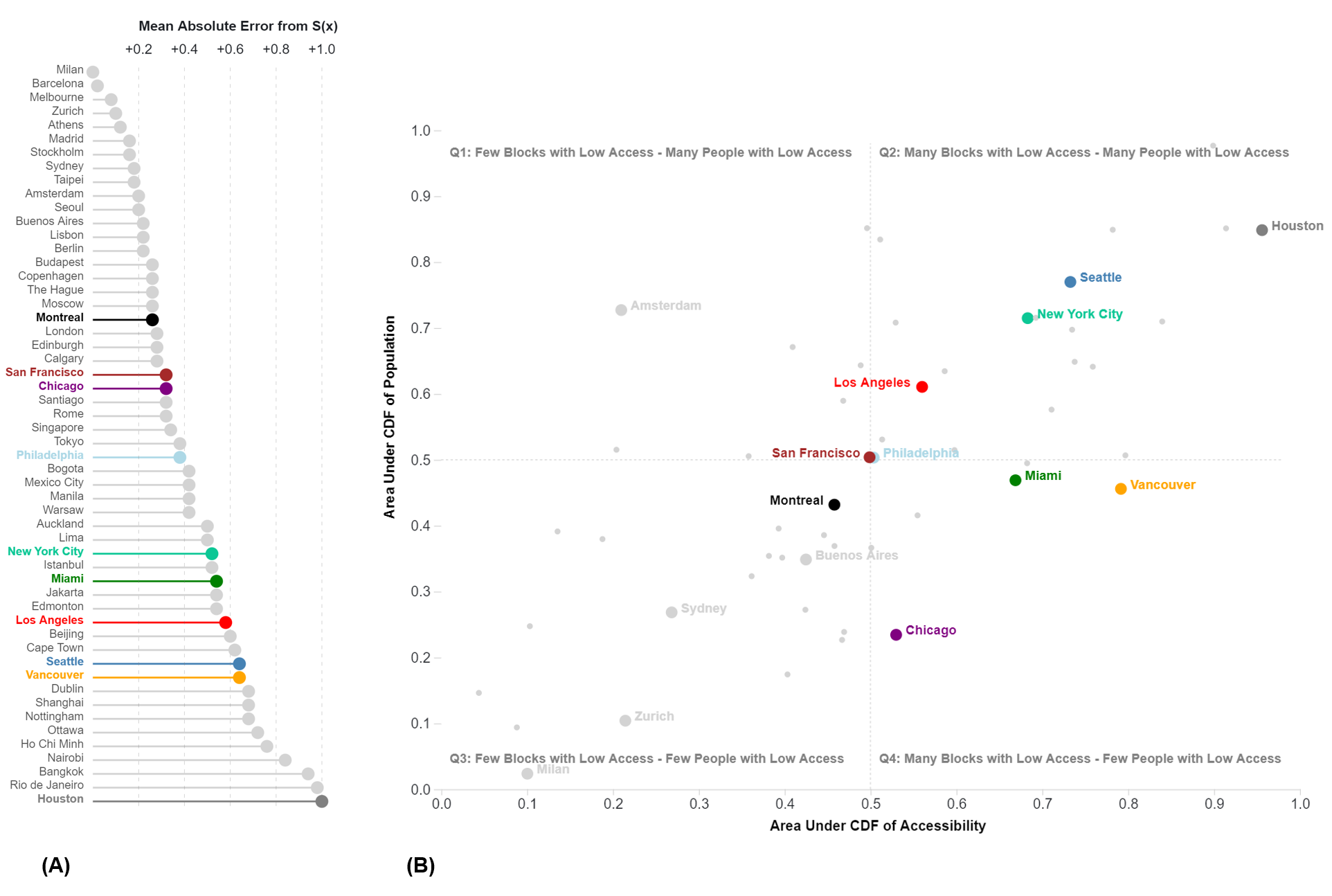}
\caption{(A) Ranking of cities using $S(x)$ as a benchmark where $50\%$ of the population has at least an accessibility score $A$ of $0.5$. Cities are ranked according to the normalized mean absolute error (MAE) between a city's population curve and the benchmark function $S(x)$. (B) The four quadrants of accessibility versus population densities. Q1 represents a scenario where a large number of spatial units have high accessibility scores but the majority of the population is concentrated in spatial units of lower accessibility. Q2 represents a scenario where a large number of spatial units have low accessibility scores and the majority of the population is concentrated in spatial units of low accessibility. Cities in Q3 provides highly accessible amenities to the most dense regions, with very limited pockets of less accessible urban areas. Quadrant Q4 has cities where both low and high density spatial units have higher levels of access, but the cities may be more sprawled with some regions where access is low.}
\label{fig:fig3}
\end{figure}

To understand this relationship better - the cumulative density (spatial units) of accessibility (CDF($A$)) and the cumulative population density of accessibility (CDF($p$)) - we present a ranking and a quadrant grid in Figure \ref{fig:fig3}. Figure \ref{fig:fig3}A represents a ranking of cities using $S(x)$ as a benchmark where $50\%$ of the population has at least an accessibility score $A$ of $0.5$. Cities are ranked according to the normalized mean absolute error (MAE) between a city's population curve and the benchmark function $S(x)$. In Figure \ref{fig:fig3}B, we plot each city's area under its (normalized) CDF($A$) and CDF($p$) against each other. Each of the four quadrants of this grid represents a unique scenario. Cities that fall in the top left quadrant (Q1) represent a scenario where a large number of spatial units have high accessibility scores but the majority of the population is concentrated in spatial units of lower accessibility. In the particular case of \textit{Amsterdam} in our data set, the city witnesses very high levels of tourism every year. Most spatial units in the city centre that have very high access, also see a huge crowd of foreign visitors \cite{stadt_ams}. Other spatial units in the city that are more residential, have lower levels of relative access within the city, but much better absolute levels of access when compared to other cities. This scenario is not observed in other cities where tourism is not so prevalent, hence the scarcity of points within this quadrant. Cities that fall in the top right quadrant (Q2) represent a scenario where a large number of spatial units have low accessibility scores and the majority of the population is concentrated in spatial units of low accessibility. This scenario is very common for the United States of America (USA) (for example, Los Angeles, Houston or Seattle) which is entrenched in policies steering away from public transportation, and a majority of the urban area is characterised by low-density suburban built environment \cite{english_why_2018}. Cities that fall in quadrant Q3 provide highly accessible amenities to the most dense regions, with very limited pockets of less accessible urban areas, but may be undervaluing areas with low density. Quadrant Q4 illustrates cities where both low and high density spatial units have higher levels of access, but the cities may be more sprawled with some regions where access is low.

\textbf{The variation in accessibility by urban profiles}

Above, we presented results on the static distribution of accessibility in any given city without differentiating among the resident population. Cities in the third quadrant of Figure \ref{fig:fig3}B have a high median accessibility, indicating that most of the population is also able to easily access urban infrastructure in their vicinity. In comparison, many cities fall in quadrant $2$, where the accessibility distribution is characterised by shorter tails and wider humps (hence more inequalities prevalent among its residents - in Ref. \cite{van2011discussing}, van Wee et. al. discuss how a Gini index measure can be used to assess levels of accessibility using density based distributions of accessibility). While this analysis of accessibility as a resource is useful, it might prevent recognition of structural inequalities in the distribution of this resource. To provide appropriate support to decision-makers in addressing socioeconomic inequalities in cities, we examine the relationship of accessibility distributions with socioeconomic attributes of households within the spatial units of a city using Chicago, IL (USA) as an illustrative example.

We cluster households using socioeconomic attributes to form urban profiles where households \emph{within} a profile are similar to each other than \emph{across} profiles (see Section Methods for description of census data and clustering methodology). We selected attributes that were largely available in the census data across Canada and USA: ethnicity and minority status, income level, marital status, household composition, language abilities, education level and type, employment status, occupation type and commuting mode. A complete list of demographic attributes is provided in the \gls{SI}. By using consensus clustering \cite{likas2003global} on this set of attributes associated with Dissemination Areas (DA) (in Canada) and Census Block Groups (CBG) (in USA) (see Section Methods for modelling details), we identified a set of urban profiles for each city in our original subset (in the Methods section we explain the choice for the subset sample for further analysis). For instance, five urban profiles are identified for the city of Chicago:
\begin{itemize}
    \item (1) Low income mixed (LIMix)
    \item (2) Low income minority (LIM)
    \item (3) Medium income white (MIW)
    \item (4) High income white (HIW)
    \item (5) Medium income white suburban (MIW-Suburban)
\end{itemize}

We observe that the resulting clusters vary between $3-5$ by city because of each city's unique socioeconomic distribution. The difference in the number of profiles for various cities is also affected by the choice of the consensus function used to evaluate the performance of clustering (see Methods for clustering performance metrics). For example, for Chicago, IL (USA) based on the k-modes consensus function, the most appropriate number of clusters is equal to 5. Importantly, for some cities the difference in the metric values is hardly distinguishable. In such cases we try to minimize the resulting number of clusters while preserving subjective cluster interpretability.

\begin{figure}[H]
\centering
\includegraphics[width = 1\textwidth]{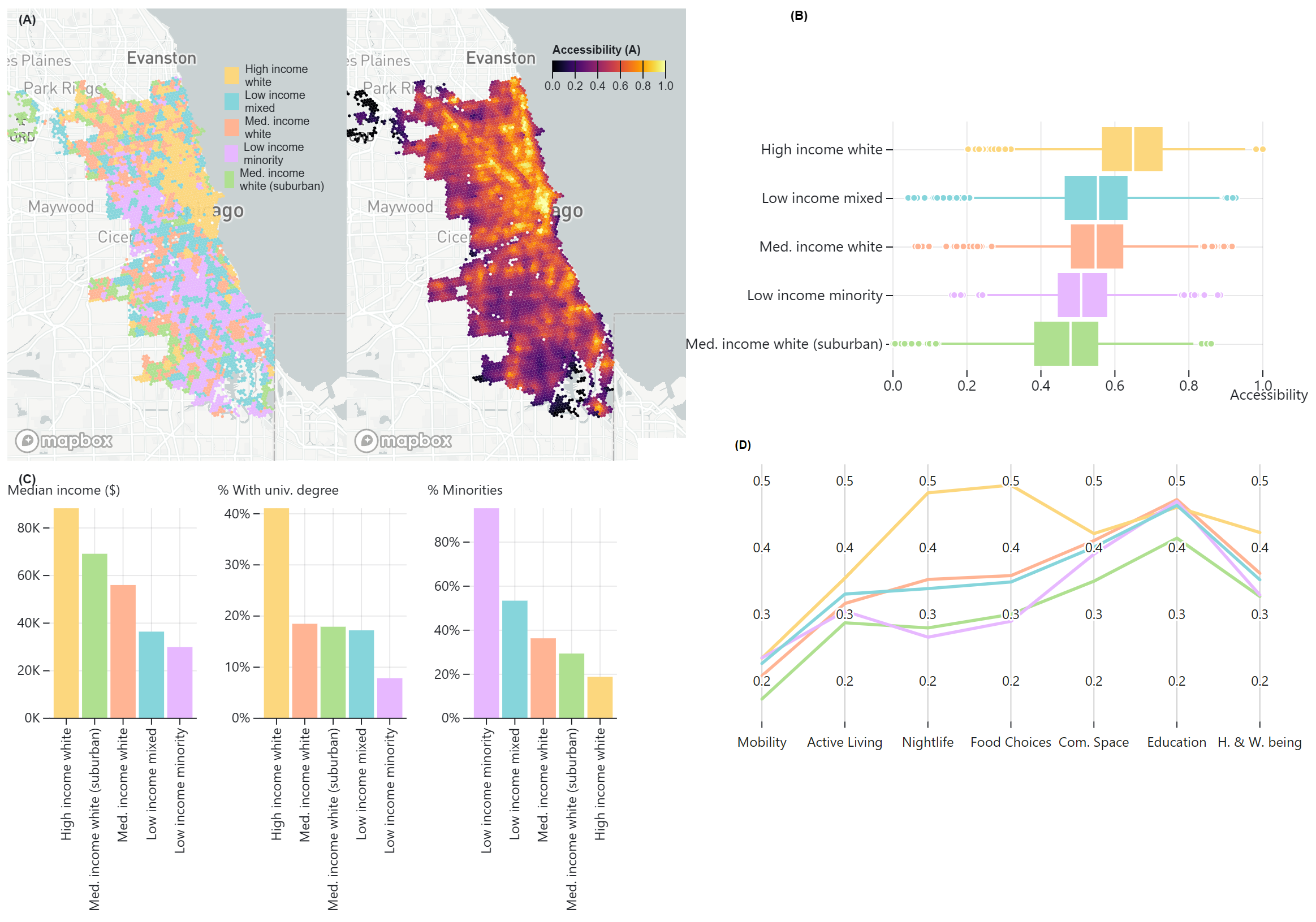}
\caption{(A) Spatial representation of the five clustered social groups for Chicago and the spatial distribution of accessibility for Chicago, where each spatial unit (an equal sized square of 0.0625 km\textsuperscript{2}) has its own accessibility score. (B) The statistical distribution of accessibility for each of Chicago's clustered social groups. (C) Statistical distribution of normalized key demographic attributes for each of Chicago's clustered social groups. (D) A parallel coordinates chart illustrating each social group's median accessibility scores dis-aggregated across amenity types.}
\label{fig:fig4}
\end{figure}

Next, to establish a relationship between the accessibility score $A$ and urban profiles, we interpolate each city's clustered DAs or CBGs onto equally sized spatial units. Although Chicago's median accessibility of 0.51 makes it generally more accessible than other cities, investigating accessibility scores for each of Chicago's urban profiles reveals more nuanced socioeconomic differences. On average, urban areas inhabited by the most disadvantaged group in Chicago are $27\%$ less accessible than those inhabited by the least disadvantaged group. Figure \ref{fig:fig4}A visually depicts the spatial differences in accessibility for each urban profile, largely indicating the disparity in accessibility between groups (2) LIM and (4) HIW, where a clear difference in contrast is visible between the areas where the two groups live. Figure \ref{fig:fig4}B illustrates the distribution of accessibility across the urban profiles in Chicago. We observe that in Chicago, urban profile (5) MIW-Suburban, represents a small fraction ($10.7\%$) of spatial units mostly located in suburban areas where low accessibility scores are likely reflective of personal lifestyle choices rather than discriminatory outcomes of public policy \cite{mikelbank2004typology}. We also observe that the (2) LIM urban profile has the least median accessibility across the other groups (sans profile (5) MIW-Suburban). The prevalent socioeconomic features of this profile are characterised by the lowest levels of income, the highest percentage of individuals who identify as visible minorities, the lowest number of individuals with university education, and the highest rates of unemployment (Figure \ref{fig:fig4}C). In contrast, the group with the highest median accessibility - (4) HIW - is in stark contrast with group (2) LIM, displaying diverging distributions for each socioeconomic attribute (Figure \ref{fig:fig4}C). If we break down accessibility by category of amenities (Figure \ref{fig:fig4}D), we observe that in Chicago, the highest degrees of inequality are found in access to infrastructure related to food choices, entertainment, active living, and health and well being. These results highlight areas for intervention, where urban planners may focus equitable accessibility policies in the city of Chicago. The results of other cities of the subset are reported in the \gls{SI}, and we can observe that in a majority of the subset of cities, low income communities who also have a larger share of minorities, earn less and are less educated, typically have lower access to urban amenities compared to other communities in the region.

\textbf{Accessibility is inequitably distributed across urban profiles}

We rank each urban profile based on their relative attribute values of income, education, minority status, and employment census variables. In each city, we choose the two groups at each extreme and classify them as either "most disadvantaged" (lowest income, lowest levels of education, highest proportion of minorities, highest unemployment) or "least disadvantaged" (highest income, highest levels of education, lowest proportion of minorities, lowest unemployment). In highly unlikely cases where the differences in socioeconomic variables are negligible between two groups, we combine them together and treat them as one. In each city, by comparing spatial units pertaining to either group with their associated accessibility scores, we calculate the probability density functions of each group's respective accessibility. The assessment of equity is determined by computing the relative change between the two groups' median accessibility as follows,

\begin{equation}
\label{delta}
\begin{split}
A_\Delta(x, y)=\frac{A_{x}-A_{y}}{A_{y}},
\end{split}
\end{equation}
where $A_{x}$ and $A_{y}$ are the respective median accessibility scores of the ``least disadvantaged" and ``most disadvantaged" groups.

\begin{figure}[H]
\centering
\includegraphics[width = 1\textwidth]{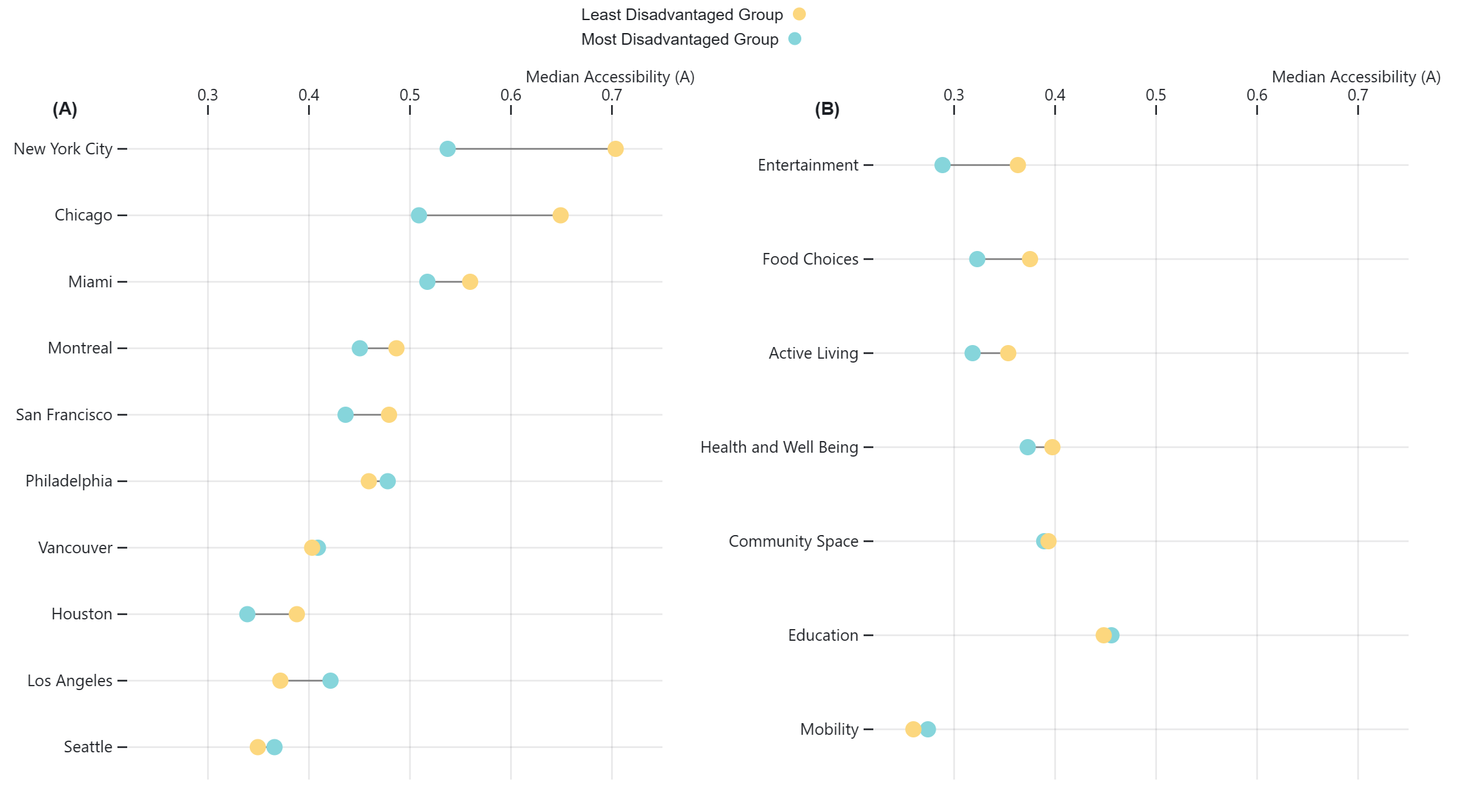}
\caption{Median accessibility scores for most advantaged and least advantaged groups aggregated by city (A) and category (B) for $10$ North American cities}
\label{fig:fig5}
\end{figure}

Figure \ref{fig:fig5} shows results from all $10$ North American cities (subset of this work). We find that inequalities in access to urban infrastructure are present in the majority ($6$ out of $10$) of cities we studied. The most disadvantaged groups are structurally under-served by urban infrastructure as compared to least disadvantaged groups. When accessibility is aggregated by category of amenities (Figure \ref{fig:fig5}), there is a gap in access to food, culture and entertainment, active living and health infrastructure, which is all more accessible in spatial units where the least vulnerable populations reside. The subset analysis illustrates that in North American cities, public transit options (represented by the mobility infrastructure) are scarcely distributed: they are either very few \cite{vey_connecting_2020} and difficult to access or have poorly evolved through complex processes of institutions, government and corporate practices that shape the choices of communities to be car-dependent \cite{english_why_2018}.

\section*{Discussion}
\label{sec:discussion}

Our goal in this paper was to investigate the spatial relationship of accessibility and urban inequalities. We found that the spatial distribution of accessibility to infrastructure follows a universal log-normal paradigm, indicating stratified urban communities (some with increasingly high access, and others with lower levels of access). In accordance with Zipf’s laws of cities \cite{gabaix1999zipf}, and considering the economic integration of urban regions \cite{cristelli2012there}, some cities exhibit a widening inequality across social and spatial lines. Our findings suggest that the distribution of amenities in cities plays a significant role in how inequalities in access to urban infrastructure manifest. The most socioeconomically disadvantaged groups are structurally under-served by urban infrastructure as compared to least disadvantaged groups. When dis-aggregating accessibility over the types of urban amenities, there is an overall gap in access to food, culture and entertainment, active living and health infrastructure, which is all more accessible in spatial units where the least vulnerable profiles reside. The societal impacts of these general inequality patterns in access to infrastructure can be serious. For example, in a study of 11,599 Philadelphia households \cite{mayer2014food}, the authors report that households who reported very difficult access to fruits and vegetables were more likely to also report food insecurity in their household. In this context, poor access to food options in communities that are already marginalized socioeconomically can contribute to perpetuate structural inequalities among urban communities. While these findings are not surprising and have been reported in multiple contexts and cases studies \cite{nijman2020urban,moro2021mobility,pereira2019desigualdades}, a framework to assess equity is an important step to understand the global state of inequalities in accessibility from a spatial perspective. Through this research we have unified previous social, empirical and theoretical work and identified the general social groups that suffer from inequitable accessibility to urban amenities.

Our research provides a holistic and reproducible framework for municipalities worldwide to redress inequalities in access to infrastructure. Because of its modular nature, the framework can be adapted to incorporate the multiple values of stakeholders with regards to accessibility, and support in identifying spatial regions and social groups suffering from inequalities in access. Starting with global case studies of countries, regions or political contexts, a lot of scholarly work can be reproduced or extended and strengthened with our framework and access to open-source data. For example, in Ref. \cite{pereira2019desigualdades}, authors concluded that white and high-income communities in Brazil have more access to jobs and education than black and poor communities, irrespective of transport options considered. Thus, by combining measures of active accessibility and dynamic measures of opportunities, such an approach can understand the big picture better, and identify pockets of poorly accessible neighbourhoods that also have lower access to job and other opportunities across an urban region. Similarly, combining our insights with work of scholars in Ref. \cite{xu2020deconstructing}, policymakers could derive targeted and equitable measures for facilitating the distribution of amenities in neighbourhoods that are most in need of such resources. Although, studying overall accessibility levels do not support particular communities in understanding their needs and contextual factors that perpetuate inequalities, our proposed approach can provide an empirical foundation to narrow down on opportunities for decision-makers to investigate further in remedying poor levels of access and inequalities. When moving to a local case study of access to particular infrastructure, this study can support scholars and policymakers in identifying situation specific factors for their neighbourhoods. This can be achieved by supplementing such a case study with other data sets that expand on in-depth socioeconomic conditions of social groups over time, and how active accessibility indicators have shaped in and around the region in conjunction.

Although this work supports in identifying general levels of accessibility for various urban communities, the results are bound to the definition of accessibility. The accessibility score we define is based on the importance attributed to each category encoded as a weight (same for all spatial units and cities; see Methods). If the weights are changed drastically, the distributions change quite considerably. Following the work of Maslow \cite{maslow1981motivation}, we agree that changing these weights drastically mean little to our understanding of the basic needs of peoples. In case where decision-makers pander to different politics where our needs change or must change considerably (and thus the weights), the associated distributions of accessibility will paint a different picture recognising the goals of policymakers as opposed to the present levels of access that are afforded to communities. As our framework design is modular and these weights can be changed, the associated distributions of accessibility can help in understanding local measures of access and contextualising them within the larger policy goals of the city itself. We propose two avenues for future work in this regard. In one, more practice-oriented work can be done to decipher the needs of communities through public participation methods. These can be translated to weights that are region specific and compared more broadly to city-wide levels of access. In two, a theoretical approach can be used carry out a sensitivity analysis (for instance, a \textit{Sobol Salteli} sampling) on the weights to evaluate the robustness of the [different] definition of accessibility. The outcome of such an analysis; results of the changes in distributions of the accessibility density plots can support in multiple policy goals where negotiations might be necessary among decision-makers and stakeholders for further development.

Urban planning varies in practice across cities around the world. Diverse and competing values, and regulatory policies around market instruments usually govern how accessibility to amenities is shaped \cite{nijman2020urban,huang2011spatial}. This is coupled with more bottom-up processes of change within neighbourhoods (effects of local economies, segregation and gentrification) that are shaped by the availability (or lack thereof) of good quality infrastructure. While all of the subset cities that we have studied have implemented policies targeted at improving equity in accessibility \cite{seskin2012complete}, the presence of such a gap in accessibility indicates that de facto processes of urban change and de jure policies (whether discriminatory government actions of the past \cite{higginbotham1990jure,parekh2017more}, or utilitarian policies and neoliberal market investments \cite{farmer2011uneven}) still perpetuate social inequalities in today's urban communities. Additionally, these results show that present accessibility policies must be more ambitious, so as to reduce the inequalities that exist in access to infrastructure to acceptable levels and improve the minimum access affordability for multiple urban communities.


\section*{Materials and Methods}
\label{sec:methods}

\subsection*{Accessibility Score}
To compute accessibility, we use the amenity location point data (from here on called Points of Interest (POI)) and the pedestrian street network data collected from OpenStreetMap, where each node represents a street intersection and each edge represents a pedestrian road segment. POIs do not warrant equal importance for all people and contribute differently to addressing inequalities \cite{klinenberg2018palaces}. For example, improving access to a pharmacy or library might be more important to a social group than access to a public transport station. Maslow's hierarchy of needs suggests, that in general, humans have three important needs in their living environment: Physiological, Safety and Security, and Social needs \cite{maslow1981motivation}. This suggests that similar POIs can be classified into categories for ease of analysis and to reflect more broadly on the concept of accessibility and inequalities. So, to capture the essential aspects of urban accessibility, we classify each POI into one of \textit{seven} categories of accessibility: Mobility, Active living, Entertainment, Food Choices, Community Space, Education, Health and Well Being (Supplementary Note 1). We choose these categories to reflect on the needs of peoples (as specified by Maslow) and broader attributes of functional and livable urban spaces: transportation, public health, food security, cultural capital, and social cohesion \cite{kenworthy2006eco,leby2010liveability}. We (the 3 co-authors) carried out a subjective assignment of the POI to the seven categories. Then we deliberated on where we differed, and found resolution within the context of needs and livable urban spaces. Next, each point, for each category, is assigned to the nearest edge on the pedestrian network. For each node on the network, we calculate the routing distance (weighted shortest path along the pedestrian network) to the nearest POI. Each node on the network is then assigned a value representing the distance in meters to the nearest amenity point. This process is repeated for each category of amenities, resulting in each node being assigned seven accessibility values. Previous studies on quantifying accessibility in cities have focused on measuring the Euclidean distance to services \cite{gastner2006optimal}. However, the ease with which residents can access services in a city is limited by urban form and land-use and the routing distance is a better proxy for the accessibility \cite{xu2020deconstructing}.

With the use of Python’s open source \emph{OSMnx} library \cite{boeing2017osmnx}, we filter OpenStreetMap according to different keys (“public transport”, “leisure”, “amenity”). We create a geospatial point data set, consisting of $50$ types of urban amenities, of all the POIs considered in each city. In total we collect $1,678,000$ POIs across all $54$ cities. \gls{SI} reports the number of points collected for each city.

Next, we use \emph{OSMnx} to download street networks for the pedestrian street infrastructure of each city. Using the street network data, we interpolate nodes within identical 0.0625 km\textsuperscript{2} spatial units in a grid matching those resulting from the analysis of social groups. By averaging distances from the nodes that fall within each spatial unit to nearest amenities of each category (that may or may not be in the same spatial unit), we compute the average walking distance within every spatial unit of a city to each of the seven categories of basic amenities. Finally, we weigh and aggregate average walking distances into a walking distance measure $D$ using the following equation:
\begin{equation}
\begin{split}
D_{i}=\log(\sum\limits_{c} w^c d_i^c),
\end{split}
\end{equation}
where $d_i^c$ is the average walking distance from nodes that fall within spatial unit \textit{i} to the nearest amenity of a given category \textit{c}, and \textit{w} (category-wise constant across all spatial units and cities) is the weight attributed to that category. See \gls{SI} for the values of weights to reproduce this analysis. The use of a logarithmic function captures the outliers in the measurement and makes comparison easier within and across cities. Using a min-max scaler function, we normalize measure $D$ within each city so we can normatively rank walking distances of every spatial unit relative to each city:


\begin{equation}
\label{norm}
\begin{split}
\widetilde{D_{i}}=\frac{D_{i}-min(D_{i})}{max(D_{i})-min(D_{i})}.
\end{split}
\end{equation}

Finally, we subtract $\widetilde{D\textsubscript{i}}$ from $1$ to obtain an accessibility score $A$, where large values indicate high accessibility and low values indicate low accessibility:

\begin{equation}
\begin{split}
A_{i}=1-\widetilde{D_{i}}
\end{split}
\end{equation}

\subsection*{Data-driven Framework}

\begin{figure}[H]
\centering
\includegraphics[width = 0.7\textwidth]{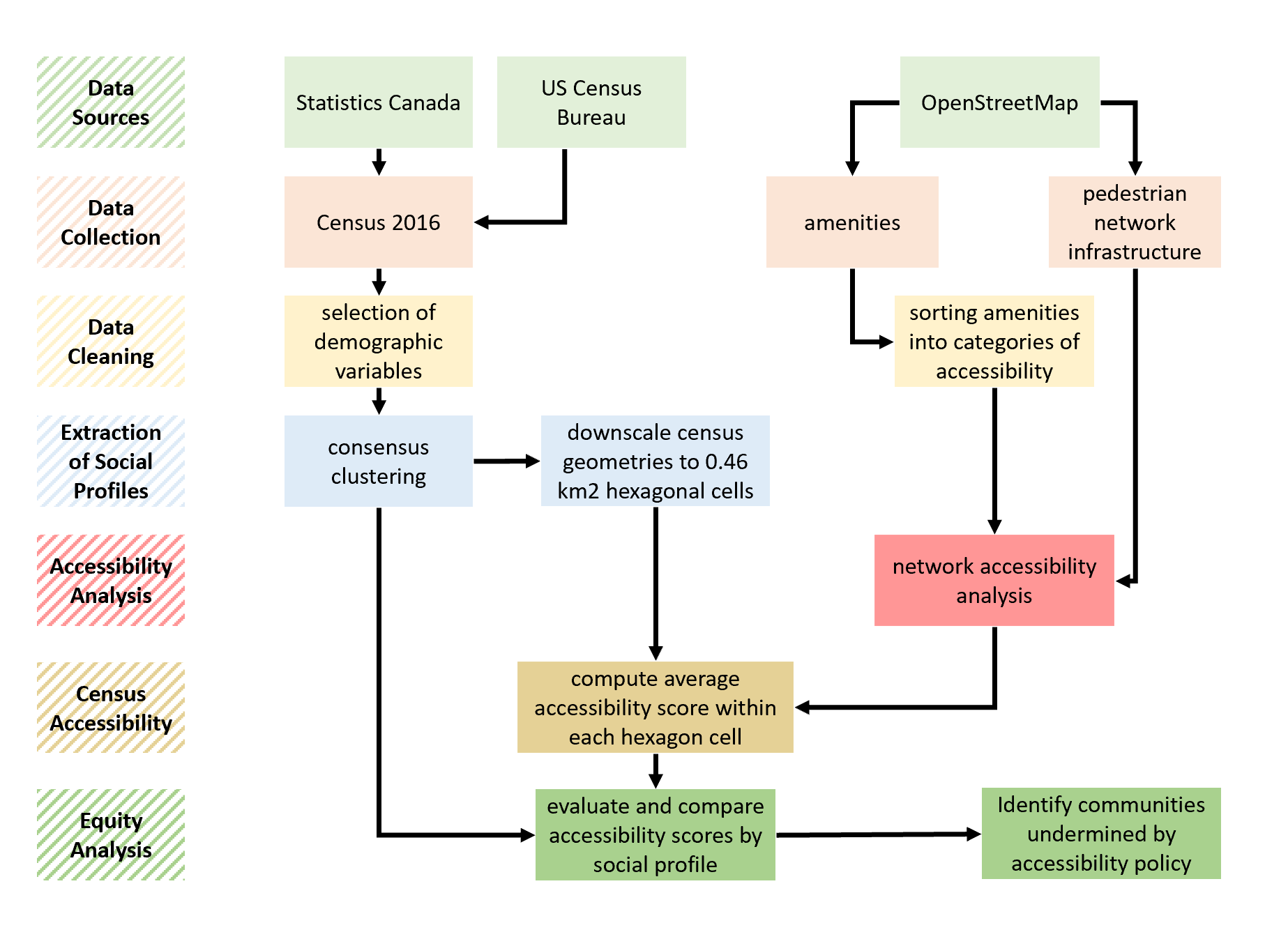}
\caption{Framework for Evaluating Equity in Accessibility to Urban Infrastructure.}
\label{fig:fig6}
\end{figure}

To understand the nature of inequalities that citizens face in accessibility to urban infrastructure, we introduce a data-driven framework that facilitates comparison of accessibility scores of groups of individuals that share similar demographic and socioeconomic attributes (e.g. age, ethnicity, income, employment, mobility) within and across cities worldwide (Figure \ref{fig:fig6}). This framework is based on three types of data sets (see Section Data Description). First, we use census data from governmental agencies to identify which population groups are predominant within a certain area - these are called \textit{social groups} (see Section Data Description for detailed information on census data). We use consensus clustering (see Section Clustering Method) to identify social groups using variables of interest. In parallel, we perform a network analysis on the street networks and urban amenities data to quantify and map accessibility to urban infrastructure at the block level. Finally, we measure inequalities by comparing the distribution of accessibility across social groups that were identified through consensus clustering. In the spirit of contributing to open-source data and research \cite{papers_with_code}, our framework is fully automated and documented as open-source code on GitHub \cite{leonardo_nicoletti_2020}.

We apply this framework to $54$ cities across \emph{six} continents as reported in the Tables in the \gls{SI}. We measure accessibility for all $54$ cities and examine the variability in its distribution for the city population on a subset of those cities: New York City (NY), Chicago (IL), San Francisco (CA), Los Angeles (CA), Houston (TX), Seattle (WA), Miami (FL), Toronto (ON), Vancouver (BC), and Montreal (QC). Within this subset, the municipal authorities of each city  have made strong policy efforts to provide equitable accessibility to urban services within their municipal jurisdiction \cite{seskin2012complete}. In addition, open-source data programs run by the municipal governments of these cities provide aggregated demographic and socioeconomic variables for the complete population.

\subsubsection*{Data Description}\label{sec:data}
\textbf{Census Data.} To characterize population demographics, we use $2016$ demographic data from Statistics Canada \cite{statistics_canada} at the Dissemination Area (DA) level, and 2017 demographic data estimates from the United States Census at the Census Block Group (CBG) \cite{us_census_bureau_2019} level. In order to capture a full spectrum of demographic attributes, we consider $46$ census variables relating to ethnicity and minority status, income level, marital status, household composition, language abilities, education level and type, employment status, occupation type and commuting mode. Several scholars agree that many of these socioeconomic variables are related to factors explaining inequalities in cities \cite{nijman2020urban}, although the evidence is scattered in the literature through case studies specific to cities or types of infrastructure.

\textbf{Streets and Amenities.} To characterize street networks and urban amenities, we use OpenStreetMap \cite{OpenStreetMap} to collect data on pedestrian infrastructure and geographically allocated places of interest (POI): hospitals, schools,  supermarkets, restaurants, schools, etc. POIs are returned by the OpenStreetMap API as amenity names and their associated geographical location (i.e. longitude and latitude). Pedestrian infrastructure networks are returned by the OpenStreetMap API as networks of nodes and edges, where each node represents a street intersection and each edge represents a segment of road with walkable or bikable features. Next, we group POI types into a set of different categories (see \gls{SI} for a complete list of amenities and their associated category).

This source is inconsistent in terms of data completeness and quality. The street network can be incomplete in several rural and developing parts of the world. The number of amenities in OSM can be twice as high as the real number (for instance, the number of POIs categorized as schools for Helsinki is around 250 whereas the real (from the city's administrative data base) is 100). In some cases, this happens because the name of buildings occur twice or thrice, like school building A, and B. In our study, we are not interested in the number of those amenities, rather, the location is important. Further, the cities we have selected have a reliable amount of data ($>80\%$) available from OpenStreetMap, and in most cases, the data is complete even in the developing world \cite{barrington2017world}.

\textbf{Population Grids}
Data on population density for every city is retrieved from the European Commission's 2015 Global Human Settlement Layer (GHSL) \cite{GHSL}. This data is retrieved in the form of a grid of 250m by 250m squares (0.0625 km\textsuperscript{2} spatial units exactly the same size as units used for accessibility analysis) and their associated population density values covering the entire world.

\textbf{Year mismatch in the collection of data sets}
The data sets we have used in this analysis have been collected and catalogued across different years. The census data is from the year 2016 (Canada) and 2017 (US). The street networks were extracted in the year 2019. The GHSL grid data of population density is from 2015. This does not cause any problems for comparison as we do not expect population densities to change drastically within a couple of years. Further, we are not making any inferences about certain people but a group as a whole, and do not expect neighbourhoods to gentrify within a year in any city in the world.

\subsection*{Clustering Method} \label{sec:clustering}
Consensus clustering is an unsupervised machine learning method that is widely used in the fields of molecular biology and complex networks \cite{monti2003consensus, kiselev2017sc3, lancichinetti2012consensus}. While it has recently found useful applications in the field of social network analysis \cite{kheirkhahzadeh2019community, kheirkhahzadeh2020consensus}, there is not much evidence for its use in urban science \cite{frias2014consensus, lopez2017revealing}. The method combines and evaluates results of several clustering algorithms to find the most robust set of labels for clusters \cite{monti2003consensus} in the given data set. The final decision about the labels is made based on a consensus function (e.g. Majority Voting \cite{ayad2010voting}). In this work, we use the k-modes consensus function \cite{luo2006combining}:
\begin{equation}
\begin{split}
d(x, y)=\sum\limits_{i=1}^m(1-\delta(x_i, y_i)),
\end{split}
\end{equation}
where \textit{x} and \textit{y} are two categorical objects (spatial units) with \textit{m} categorical attributes (demographic and socioeconomic variables), and the total mismatches of the corresponding attribute categories of the two objects represent the dissimilarity measure between x and y \cite{luo2006combining}.

Authors of Ref. \cite{luo2006combining} showed that such an approach outperforms \textit{other} consensus functions and has low computational complexity. However, since the true labels are unknown, we have not evaluated results of other consensus functions versus k-modes. In future endeavours, it might be interesting to validate the clusters with expert knowledge of the urban region, and compare that with other consensus functions.

We manually investigated over 16,000 socioeconomic variables for the US and over 4,000 for Canada and selected 44 variables of two categories of interest: socioeconomic (age, ethnicity, income, employment) and housing-related (type of house ownership, average housing price). The number of observations (the sample) differ for each of the cities. All features are numerical with most of them reported as percentages. The resulting data set has outliers. There are areas in cities where, for example, population density is significantly smaller than in the rest of the blocks. To include all areas in the analysis appropriately, we reduce the influence of outliers through a technique called winsorizing \cite{wilcox2005trimming}: limiting the number of extreme values in the data set. We set up the limits as 0.1 percentile and 0.9 percentile respectively, hence subjecting the data to 80\% winsorization. In the final pre-processing step, we standardize all features by removing the mean and scaling the observations to unit variance.

To implement consensus clustering in our analysis we use the \textit{diceR} library for the R programming language, which uses an ensemble clustering framework to perform cluster analysis \cite{chiu2018dicer}. We iterate over 5 clustering algorithms, namely K-means, K-medoids, Hierarchical clustering, Gaussian mixture model and HDBSCAN. diceR also requires to specify distance functions. We select 7 distance functions aiming to cover all potential issues with the data: Euclidean, Maxmium, Manhattan, Canberra, Binary, Minkowski and Spearman. The algorithm runs 10 replications and gives us cluster labels based on k-modes consensus function.

\subsection*{Spatial Units}
In this study, each spatial unit is described by a 0.0625 km\textsuperscript{2} square originating from the GHSL population grid \cite{GHSL}. The main spatial unit is a square that does not delineate the spatial context on the ground (data observation or public administration of space as differently sized and populated administrative districts). We use square grids to reduce sampling bias from edge effects \cite{mayaud2019future} and minimize the Modifiable Areal Unit Error (MAUP) \cite{dark2007modifiable}. Although this framing does resolve the MAUP partially, it does not represent a spatial unit that is directly linked to the problem at hand, and thus introduces MAUP on its own. Like any spatial unit that brings in a degree of MAUP, squares and hexagons have the ability to partly address the size problem. If the data is also collected keeping in mind the framed boundaries of equally-spaced grid-cells, a better resolution could be achieved. However, this is not the case with population and accessibility variables, and interpolation of these data sets is bound to bring in some error. For each city, the GHSL grid is clipped to the city's administrative boundaries and is used to calculate the accessibility score (A) within each of its squares. For each of the 10 cities in the subset analysis of socioeconomic variables, this same grid is used to assign cluster values obtained from clustering using census geometries. This process involves (a) checking whether a given square spatially intersects with a given census geometry, (b) assigning this census geometry's cluster value to the given square if this condition is met. Note that our methodology here is limiting in nature. Each census geometry is assigned one cluster label from the consensus clustering algorithm. The square spatial units that fall entirely within the census geometries get one clearly defined label attributed to it. However, for squares that fall at the boundary of census geometries \textit{and} the boundary of two clusters ($<2\%$ in the entire data set), their assignment to a cluster is based only on the order in which they are being checked.

\end{document}